\documentclass[twocolumn]{aastex7}
\usepackage{amsmath}
\usepackage{amssymb}
\usepackage{graphicx}
\usepackage{booktabs}
\graphicspath{{./}}

\begin{document}

\title{Migration Traps as Variability Attractors. II. Nonmonotonic X-Ray Timing Signatures of Embedded Stellar-Mass Black Holes in AGN Disks}

\shorttitle{X-Ray Timing Signatures of Embedded sBHs}
\shortauthors{Xing et al.}

\correspondingauthor{Tong Liu}
\email{tongliu@xmu.edu.cn}

\author[0009-0008-9726-9431]{Jing-Tong Xing}
\affiliation{Department of Astronomy, Xiamen University, Xiamen, Fujian 361005, China}
\email{kriseme@163.com}

\author[0000-0001-8678-6291]{Tong Liu}
\affiliation{Department of Astronomy, Xiamen University, Xiamen, Fujian 361005, China}
\affiliation{SHAO-XMU Joint Center for Astrophysics, Xiamen University, Xiamen, Fujian 361005, China}
\email{tongliu@xmu.edu.cn}

\author[0000-0002-0771-2153]{Mouyuan Sun}
\affiliation{Department of Astronomy, Xiamen University, Xiamen, Fujian 361005, China}
\email{msun88@xmu.edu.cn}

\author[0000-0002-4223-2198]{Zhen-Yi Cai}
\affiliation{Department of Astronomy, University of Science and Technology of China, Hefei, Anhui 230026, China}
\affiliation{School of Astronomy and Space Science, University of Science and Technology of China, Hefei, Anhui 230026, China}
\email{zcai@ustc.edu.cn}

\author[0000-0002-7329-9344]{Ya-Ping Li}
\affiliation{Shanghai Astronomical Observatory, Chinese Academy of Sciences, Shanghai 200030, China}
\affiliation{SHAO-XMU Joint Center for Astrophysics, Xiamen University, Xiamen, Fujian 361005, China}
\email{liyp@shao.ac.cn}

\author[0009-0005-2801-6594]{Shuying Zhou}
\affiliation{Department of Astronomy, Xiamen University, Xiamen, Fujian 361005, China}
\email{zhoushuying@stu.xmu.edu.cn}

\author[0000-0003-1474-293X]{Da-Bin Lin}
\affiliation{Laboratory for Relativistic Astrophysics, Department of Physics, Guangxi University, Nanning 530004, China}
\email{lindabin@gxu.edu.cn}

\author[0000-0001-9449-9268]{Jian-Min Wang}
\affiliation{Key Laboratory for Particle Astrophysics, Institute of High Energy Physics, Chinese Academy of Sciences, Beijing 100049, China}
\affiliation{School of Astronomy and Space Sciences, University of Chinese Academy of Sciences, Beijing 100049, China}
\affiliation{National Astronomical Observatory of China, Beijing 100020, China}
\email{wangjm@ihep.ac.cn}

\begin{abstract}
Stellar-mass black holes (sBHs) embedded in active galactic nucleus (AGN) disks may affect X-ray variability through magnetic reconnection. As sBHs migrate and accumulate near migration traps, they disturb the disk magnetic field and trigger reconnection events that heat the disk and eject plasma into the overlying corona. The injected plasma dissipates kinetic energy and can trigger the release of magnetic energy already stored in the corona. We calculate the resulting coronal X-ray emission and the disk ultraviolet (UV) variability, including the disk thermal response and reprocessing of incident X-rays. We find that embedded sBHs can initially enhance short-timescale X-ray variability, but the variability becomes suppressed as more sBHs accumulate in the disk. At higher mass accretion rates through AGN disks, this suppression appears earlier. As sBH-associated reconnection events become more frequent, their active regions increasingly overlap, reducing the magnetic energy available to each event from the corona. We also find that variations in the disk ultraviolet (UV) emission follow the X-ray variations with a shorter delay than in the model without sBHs. These results suggest that a growing population of embedded sBHs does not necessarily produce stronger short-timescale X-ray variability.
\end{abstract}


\section{Introduction}
\label{sec:introduction}

Active galactic nuclei (AGNs) show flux variability in the X-ray, ultraviolet (UV), and optical bands. Their X-ray emission is generally thought to originate from the hot corona above the inner accretion disk. Inverse Compton scattering in the corona can produce hard X-ray spectra \citep[e.g.,][]{1991ApJ...380L..51H,1993ApJ...413..507H,2015MNRAS.451.4375F,2018MNRAS.480.1819R}. Power spectral densities (PSDs) and structure functions (SFs) are commonly used to study X-ray variability on different timescales \citep[e.g.,][]{2003ApJ...593...96M,2003MNRAS.345.1271V,2006Natur.444..730M}. Multiwavelength monitoring has also revealed correlations and time lags between the X-ray and UV/optical light curves. These lags may be affected by irradiation of the accretion disk and the thermal response of the disk gas \citep[e.g.,][]{2015ApJ...806..129E,2016ApJ...821...56F,2017ApJ...835...65S,2017ApJ...840...41E,2018MNRAS.480.2881M}. However, the physical origin of some AGN variability remains unclear.

Stars and compact objects can be captured by AGN disks and migrate under the action of disk torques \citep[e.g.,][]{1991MNRAS.250..505S,1993ApJ...409..592A,2007MNRAS.374..515L}. When the net disk torque changes sign at a certain radius, inwardly and outwardly migrating objects can accumulate near a migration trap \citep[e.g.,][]{2016ApJ...819L..17B,2019ApJ...878...85S}. Embedded stellar-mass black holes (sBHs) may continue to accrete, interact, and form binaries in these regions. Previous studies have mainly focused on black hole mergers and gravitational-wave sources in AGN disks \citep[e.g.,][]{2012MNRAS.425..460M,2017ApJ...835..165B,2017MNRAS.464..946S,2019PhRvL.123r1101Y,2020ApJ...898...25T,2021ApJ...911..124L}. Recent work has also considered the effects of embedded objects on the persistent electromagnetic emission of AGNs \citep[e.g.,][]{2023ApJ...946L...3T,2025MNRAS.537.3396E,2024ApJ...966L...9Z,2025ApJ...991..167X}.

Magnetic reconnection may occur in both AGN disks and coronae. An embedded sBH can disturb the surrounding gas and change the local magnetic-field structure, thereby triggering reconnection within the disk \citep{2025ApJ...991..167X}. Part of the released energy heats the disk, while some highly magnetized plasma may escape into the corona. The escaping plasma can disturb the existing coronal magnetic loops and cause further reconnection. Thus, the energy released in the corona may come from both the kinetic energy of the incoming plasma and the magnetic energy already stored in the corona \citep[e.g.,][]{2008ApJ...682..608U,2008ApJ...688..555G}.

In Paper~I \citep{2026ApJ..1010..134X}, we studied the evolution of sBH populations in AGN disks and their effects on optical/UV variability. We calculated the disk magnetic reconnection events associated with the sBHs and included the effects of local heating and gas depletion on the disk. In this paper, we investigate the corresponding X-ray emission. We use the sBH populations and reconnection events obtained in Paper~I to calculate the plasma energy entering the corona, the replenishment and release of coronal magnetic energy, and the resulting X-ray light curves. We then study their PSDs, SFs, and UV--X-ray lags for different population ages and AGN disk accretion rates.

Section~\ref{sec:model} describes the disk reconnection events, coronal magnetic-energy model, and calculation of the X-ray and UV light curves. Section~\ref{sec:results} presents the variability and time-lag results. Their physical interpretation and observational implications are discussed in Section~\ref{sec:discussion}. The main conclusions are summarized in Section~\ref{sec:conclusions}.

\section{Physical Model}
\label{sec:model}

We investigate how plasma ejected from an AGN disk affects X-ray variability after entering the corona. The embedded sBHs migrate through the AGN disk and may accumulate near migration traps. We adopt the resulting radial distributions of their number and mean mass, together with the associated gas depletion, as functions of population age. The embedded sBHs disturb the local magnetic field and produce disk reconnection events \citep{2026ApJ..1010..134X}, which heat the disk and eject plasma into the corona. We first determine the events that eject plasma from the disk and then calculate the interaction between the plasma and the coronal magnetic field. A fraction of the dissipated energy is converted into X-ray emission, from which we calculate the light curves. The energy retained in the disk also contributes to the UV emission through local heating. We prescribe the storage and release of coronal magnetic energy without solving the three-dimensional magnetic-field structure or full radiative transfer self-consistently.

\subsection{Embedded sBHs and Magnetic Reconnection in the Disk}
\label{subsec:disk_population}

We adopt a steady $\alpha$-disk model \citep[e.g.,][]{1973A&A....24..337S,1981ARA&A..19..137P}. The supermassive black hole (SMBH) mass is $M_\bullet=5\times10^7\,M_\odot$, and the viscosity parameter is $\alpha=0.2$. We consider dimensionless accretion rates $\dot m=\dot M/\dot M_{\rm Edd}=0.02$, 0.05, and 0.20, where $\dot M_{\rm Edd}=10L_{\rm Edd}/c^2$. We define the gravitational radius as $R_g=GM_\bullet/c^2$, the Schwarzschild radius as $R_{\rm S}=2R_g$, and the Keplerian angular frequency as $\Omega_{\rm K}=(GM_\bullet/R^3)^{1/2}$. For a nonspinning SMBH, the innermost stable circular orbit is $R_{\rm ISCO}=6R_g=3R_{\rm S}$. We divide the radial range $10R_{\rm S}\leq R\leq5000R_{\rm S}$ into 200 logarithmically spaced annuli. At each radius, we use the disk midplane temperature $T_{\rm c}$, surface density $\Sigma$, vertically integrated pressure $\Pi$, and scale height $H$. The two-sided viscous heating rate is $Q_{\rm vis}^0=3GM_\bullet\dot M f_R/(4\pi R^3)$, with $f_R=1-(R_{\rm ISCO}/R)^{1/2}$. As in Paper~I, we adopt a mean magnetic-coupling heating rate $Q_{\rm mc}^0=Q_{\rm vis}^0/3$ \citep[e.g.,][]{2001MNRAS.321..549M,2020ApJ...891..178S}.

For each accretion rate, Paper~I provides the sBH number $N_{\rm sBH}(R,t_{\rm pop})$, mean mass $m_{\rm sBH}(R,t_{\rm pop})$, and gas-depletion factor $f_{\rm gap}(R,t_{\rm pop})$ as functions of radius and population age. We use three population ages: $t_{\rm pop}=10^2$, $5\times10^4$, and $5\times10^5$~yr. The local gas density and surface density are multiplied by $f_{\rm gap}$, with a lower limit of 0.5, following Paper~I \citep[e.g.,][]{2006Icar..181..587C,2018ApJ...861..140K}. We set $f_{\rm gap}=1$ in the reference model without embedded sBHs.

We adopt the disk-side magnetic reconnection model developed in Paper I, in which reconnection occurs in local current sheets associated with embedded sBHs. Here we use the same reconnection prescription, including the ram-pressure factor $g_{\rm ram}(R)$, the event duration, and the energy released in each event. The event rate in an annulus is
\begin{equation}
\dot N_{\rm I}(R,t_{\rm pop}) =
\frac{N_{\rm sBH}(R,t_{\rm pop})\,g_{\rm ram}(R)}
{t_{\rm A,d}(R)} ,
\label{eq1}
\end{equation}
where $t_{\rm A,d}=H/v_{\rm A,d}$ is the Alfv\'en crossing time in the disk, $v_{\rm A,d}$ is the local Alfv\'en speed, and $0\leq g_{\rm ram}\leq1$. We take the current-sheet length to be $\ell_{\rm sh}\simeq20R_{\rm S,sBH}$, where $R_{\rm S,sBH}=2Gm_{\rm sBH}/c^2$. Its thickness is $\delta_{\rm sh}=g_s\ell_{\rm sh}$, with $g_s=0.12$. The reconnection duration is $t_{\rm I}\simeq\delta_{\rm sh}/(0.1v_{\rm A,d})$. The energy of a disk reconnection event is estimated as $E_I=L_{I,\rm pk}t_I$, where the peak luminosity $L_{I,\rm pk}$ is obtained by integrating the annulus-averaged peak heating flux given in Equation (31) of Paper~I.

The mass of an upward-moving plasma packet is $m_{\rm pkt}=\rho_{\rm d}A_{\rm pkt}\delta_{\rm sh}$, where $\rho_{\rm d}$ is the disk density after gas depletion. We approximate the upward-moving plasma packet as a column with a circular cross section of diameter $\ell_{\rm sh}$. Its projected area is therefore $A_{\rm pkt}=\pi(\ell_{\rm sh}/2)^2$. This is a simplified geometric prescription for the plasma ejected by a disk reconnection event. The reference collision area of one packet is $A_{\rm coll}=\min(A_{\rm pkt},A_{\rm ring})$. The annular area $A_{\rm ring}=2\pi R\Delta R$ provides an upper limit to the projected collision area; it does not imply that a single plasma packet interacts with the entire annulus. We assume that a fraction $\eta_{\rm up}=0.20$ of the energy released by an event is carried upward as plasma kinetic energy. We adopt this fraction as a fiducial value, and the initial kinetic energy carried by an event with stochastic amplitude is $K_{\rm up}=\eta_{\rm up}{\cal A}_{\rm I}E_{\rm I}$. The remaining energy, $(1-\eta_{\rm up}){\cal A}_{\rm I}E_{\rm I}$, stays in the disk and contributes to local heating. Here ${\cal A}_{\rm I}$ is a dimensionless random multiplier that describes the variation in the strength of individual reconnection events. Its statistical distribution is specified below.

For a timestep $\Delta t$, the event number follows a Poisson distribution with mean $\dot N_{\rm I}\Delta t$. When the expected count exceeds 30, we use a Gaussian approximation for the event number and a moment-matched lognormal approximation for the summed amplitude. The amplitude of a single event is ${\cal A}_{\rm I}=\exp(\sigma_{\rm I}Z-\sigma_{\rm I}^2/2)$, where $Z$ is a standard normal variate and $\sigma_{\rm I}=0.8$. These events provide the time-dependent plasma input to the corona. We calculate the mass, energy, and arrival time of the plasma from each annulus.

\subsection{Plasma Injection and Magnetic Reconnection in the Corona}
\label{subsec:stochastic_background}

\begin{figure*}[t]
\centering
\includegraphics[width=0.8\textwidth]{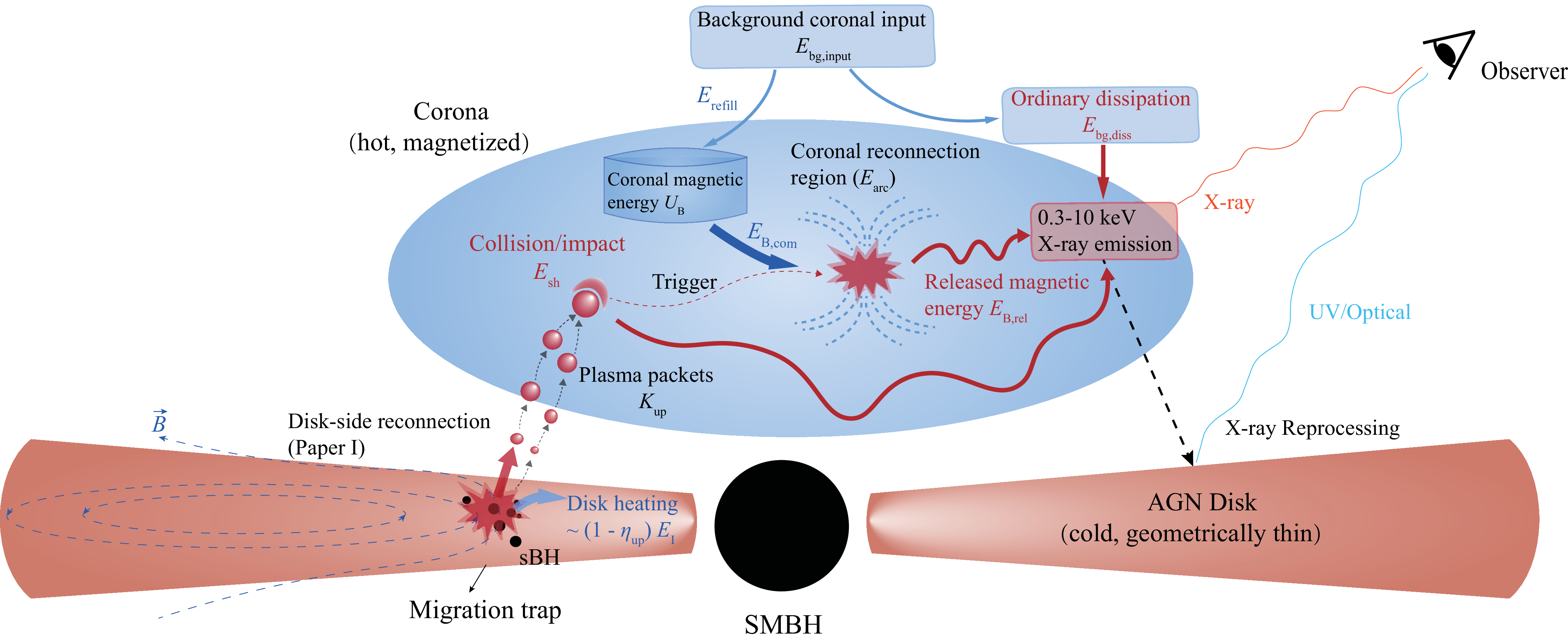}
\caption{Schematic diagram of the disk-corona energy transfer in our model. Magnetic reconnection associated with embedded sBHs ejects plasma from the AGN disk, while part of the released energy remains in the disk and contributes to local heating. The ejected plasma carries kinetic energy $K_{\rm up}$ into the corona, where collisions with coronal gas dissipate energy $E_{\rm sh}$ and may trigger further magnetic reconnection. The ordinary coronal energy input $E_{\rm bg,input}$ is divided between replenishing the stored magnetic energy $U_B$ and ordinary dissipation $E_{\rm bg,diss}$. Magnetic reconnection triggered by plasma impact transfers magnetic energy $E_{B,\rm com}$ from $U_B$ to $E_{\rm arc}$, which is gradually released as $E_{B,\rm rel}$. A fraction $\eta_{\rm X}$ of the dissipated energy, $E_{\rm bg,diss}+E_{\rm sh}+E_{B,\rm rel}$, contributes to the 0.3--10 keV X-ray emission. Part of the X-ray emission irradiates the disk and contributes to the UV/optical response.}
\label{fig1}
\end{figure*}

We assume that part of the accretion power is stored in the magnetic field of the corona. We distinguish three energy components: the ordinary coronal input $E_{\rm bg,input}$, the energy $E_{\rm sh}$ dissipated when incoming plasma collides with coronal gas, and the magnetic energy $U_B$ already stored in the corona. The physical model is shown in Figure~\ref{fig1}. The incoming plasma dissipates part of its kinetic energy in the collision and can trigger reconnection of the existing coronal magnetic field. In addition to the plasma injected by disk reconnection events, ordinary accretion activity continuously supplies energy to the corona. We represent this accretion-powered background input using the reference X-ray luminosity $L_{\rm X,0}(t)$, whose mean normalization is tied to the disk accretion power and whose fluctuations are described by the stochastic process below. This input is divided between ordinary coronal dissipation and replenishment of the stored magnetic energy. Both the plasma collision and the release of coronal magnetic energy contribute to the X-ray emission.

We prescribe the coronal magnetic pressure as $P_B=Q_{\rm mc}^0/(2H\Omega_{\rm K})$ and set the coronal height to $H_{\rm cor}=H$. The maximum magnetic energy stored in an annulus is $U_{B,\rm cap}=P_BA_{\rm ring}H_{\rm cor}$, and the stored energy satisfies $0\leq U_B\leq U_{B,\rm cap}$. In this prescription, the magnetic pressure defines a fixed reference storage capacity, while the actual stored magnetic energy evolves through replenishment and reconnection. We do not explicitly model the time evolution of the coronal magnetic field. We define the magnetic-energy filling fraction before reconnection as $f_{\rm res}=U_B/U_{B,\rm cap}$. We also define the global magnetic-energy deficit as $\chi_{\rm dep}=\langle\sum_R(U_{B,\rm cap}-U_B)/\sum_RU_{B,\rm cap}\rangle_t$, averaged over time after magnetic energy has been released. These quantities indicate how much magnetic energy remains available as the number of sBH events increases.

We model the stochastic variations of the ordinary coronal power with a stationary Ornstein-Uhlenbeck (OU) process. This is equivalent to a first-order continuous-time autoregressive process \citep[e.g.,][]{2009ApJ...698..895K,2014ApJ...788...33K}:
\begin{equation}
d\xi(t) = -\frac{\xi(t)}{\tau}\,dt +\sqrt{\frac{2}{\tau}}\,dW(t), 
\label{eq2}
\end{equation}
where $W(t)$ is a Wiener process. This equation ensures that the random variable $\xi(t)$ has unit stationary variance, and its autocorrelation properties are determined by the damping time $\tau$. The damping time at radius $R$ is $\tau_R=\tau_0(R/R_{\rm cor,in})^{\alpha_\tau}$, with $\tau_0=6$~days and $\alpha_\tau=1$. To describe variations at different radii, we use two components. The first is a correlated component $\xi_{{\rm c},R}$, obtained by locally smoothing a random sequence shared by all annuli. The second is an independent component $\xi_{{\rm i},R}$, generated separately for each annulus. Both are applied only within the active coronal region, $R_{\rm cor,in}\simeq36.7R_g\leq R\leq R_{\rm cor,out}=2R_{\rm cor,in}\simeq73.4R_g$. Plasma ejected outside this region is also not included in the additional coronal X-ray emission. The combined fluctuations are
\begin{equation}
X_R(t) = \sigma_{\rm c}(R)\,\xi_{{\rm c},R}(t)
+\sigma_{\rm i}\,\xi_{{\rm i},R}(t),
\label{eq3}
\end{equation}
where the radial amplitude is
\begin{equation}
    \sigma_{\rm c}(R)=\sigma_{\rm c,0}\left(\frac{R}{R_{\rm cor,in}}\right)^{-1/2}.
    \label{eq4}
\end{equation}
We adopt $\sigma_{\rm c,0}=0.18$ and $\sigma_{\rm i}=0.22$ as fiducial amplitudes of the radially correlated and independent stochastic fluctuations, respectively. These parameters control the relative strengths of the two fluctuation components and are held fixed when comparing different sBH populations. They specify the reference stochastic driving rather than a fit to the PSD of an individual AGN. The reference X-ray luminosity without embedded sBHs is
\begin{equation}
L_{\rm X,0}(t) = \overline{L}_{\rm X,0}
\frac{\sum_{R_{\rm cor,in}\le R\le R_{\rm cor,out}} w_R\exp[p_{\rm X}X_R(t)]}
{\left\langle\sum_{R_{\rm cor,in}\le R\le R_{\rm cor,out}}
w_R\exp[p_{\rm X}X_R(t)]\right\rangle_t} ,
\label{eq5}
\end{equation}
where $\sum_Rw_R=1$, and $w_R\propto A_{\rm ring}(Q_{\rm vis}^0+Q_{\rm mc}^0)$. We adopt an exponential response to the stochastic fluctuations, following the observed rms-flux relation in accreting black holes \cite[e.g.,][]{2005MNRAS.359..345U}. The parameter $p_{\rm X}=4$ controls the variability amplitude.  The normalization is $\overline L_{\rm X,0}=C_X\sum_{R\leq7R_{\rm in}}A_{\rm ring}(Q_{\rm vis}^0+Q_{\rm mc}^0)$, where $R_{\rm in}=10R_{\rm S}$ and $C_X=0.075$. This normalization is motivated by the observed fraction of AGN radiation emitted in X-rays \citep[e.g.,][]{2020A&A...636A..73D}. Equation~(\ref{eq5}) specifies the reference coronal power. The OU process provides the same stochastic input for the different models. It is not intended to fit the full observed X-ray PSD of AGNs.

The ordinary coronal input energy in each annulus and timestep is $E_{\rm bg,input}=L_{\rm X,0}(t)\Delta t\,w_R/\eta_{\rm X}$. We adopt $\eta_{\rm X}=0.15$ as a fiducial fraction of dissipated energy assigned to the modeled 0.3--10~keV X-ray band. The energy used to replenish the coronal magnetic field is $E_{\rm refill}=\min(U_{B,\rm cap}-U_B,E_{\rm bg,input})$. Any input energy that exceeds the available magnetic-energy capacity undergoes ordinary dissipation rather than being added to $U_B$. The remaining energy undergoes ordinary dissipation, $E_{\rm bg,diss}=E_{\rm bg,input}-E_{\rm refill}$. After replenishment, the stored magnetic energy is $U_B^+=\min(U_{B,\rm cap},U_B+E_{\rm refill})$. We use this updated energy when calculating the energy released by a new reconnection event. If the magnetic energy has already reached its maximum and no new event is triggered, the ordinary input is dissipated rather than stored in the magnetic field.

We next calculate the collision between the incoming plasma and the coronal gas. The mass of the coronal column struck by a packet is $m_{\rm c}=\rho_{\rm cor}A_{\rm coll}H_{\rm cor}$. We adopt a Thomson optical depth $\tau_{\rm T}=1.5$ \cite[e.g.,][]{2021A&A...655A..60A}. The coronal density is then $\rho_{\rm cor}=\tau_{\rm T}m_{\rm p}/(\sigma_{\rm T}H_{\rm cor})$, where $m_{\rm p}$ is the proton mass and $\sigma_{\rm T}$ is the Thomson cross section. Assuming a perfectly inelastic collision, the energy converted into internal energy follows from four-momentum conservation:
\begin{equation}
E_{\rm sh} = \left[(m_{\rm pkt}+m_{\rm c})^2c^4+2m_{\rm c}c^2 K_{\rm up}\right]^{1/2}
-(m_{\rm pkt}+m_{\rm c})c^2.
\label{eq6}
\end{equation}
The difference between the initial plasma kinetic energy and $E_{\rm sh}$ remains in bulk motion and is not included in the X-ray energy input. In the numerical calculation, we combine packets that arrive in the same annulus and timestep before evaluating the collision. For a group of arrivals $b$, we define $S_b=\sum_{k\in b}{\cal A}_{{\rm I},k}$. The total kinetic energy and mass are then $K_{\rm up,b}=\eta_{\rm up}S_bE_{\rm I}$ and $m_{\rm pkt,b}=S_bm_{\rm pkt}$. The effective collision area is $A_{\rm coll,b}=\min(S_bA_{\rm coll},A_{\rm ring})$, and the mass of the coronal gas involved is $m_{{\rm c},b}=\rho_{\rm cor}A_{\rm coll,b}H_{\rm cor}$, with $A_{\rm ring}$ serving as the maximum allowed projected area. This corresponds to approximating the incoming packets as having the same characteristic column mass per unit area. Substituting these values into Equation~(\ref{eq6}) gives $E_{\rm sh,b}$.

The incoming plasma can disturb the existing magnetic field and form a local current sheet \citep[e.g.,][]{2008ApJ...688..555G,2008ApJ...682..608U,2020ApJ...903...95H,2009PhPl...16k2102B}. We assume that magnetic reconnection is triggered when
\begin{equation}
\Theta_b =
\frac{E_{\rm sh,b}+\alpha_{\rm U} U_{B,\rm coll,b}}
{U_{B,\rm coll,b}}
\geq\Theta_{\rm crit},
\label{eq7}
\end{equation}
where $U_{B,\rm coll,b}=P_BA_{\rm coll,b}H_{\rm cor}$ is the magnetic energy in the impacted coronal column. We adopt $\Theta_{\rm crit}=0.30$ and $\alpha_{\rm U}=0.20$ in the fiducial model, which is equivalent to $E_{\rm sh,b}/U_{B,\rm coll,b}\geq0.10$. These parameters describe the assumed onset condition rather than a universal threshold for magnetic reconnection.

We adopt fast magnetic reconnection in fragmented current sheets at high Lundquist numbers \citep[e.g.,][]{2009PhPl...16k2102B,2009PhRvL.103f5004D,2017ApJ...850..142C}. The reconnection inflow speed is $v_{\rm in}=\epsilon_{\rm rec}v_{\rm A}$, where $\epsilon_{\rm rec}=0.03$ \citep[e.g.,][]{2010PhRvL.105w5002U,2012PhPl...19d2303L}. The coronal Alfv\'en speed is $v_{\rm A}=c[\sigma_{\rm cor}/(1+\sigma_{\rm cor})]^{1/2}$, where $\sigma_{\rm cor}=2P_B/(\rho_{\rm cor}c^2)$ is the magnetization parameter. The local reconnection time is $t_{\rm rec}=H_{\rm cor}/(\epsilon_{\rm rec}v_{\rm A})$. We assume that reconnection spreads along the connected magnetic fields at the Alfv\'en speed \citep[e.g.,][]{2012JGRA..11710101S}. The spreading length is $\ell_{\rm sp}=\min(2v_{\rm A}t_{\rm rec},2\pi R)$. The area of one magnetic loop is $A_{\rm loop}=\min(\pi H_{\rm cor}^2,A_{\rm ring})$. The effective area affected by one event is
\begin{equation}
    A_{\rm arc} = \min\!\left[\max\!\left(H_{\rm cor}\ell_{\rm sp},\,A_{\rm loop}\right),\,A_{\rm ring}\right].
    \label{eq8}
\end{equation}
The plasma reaches the corona after $t_{\rm rise}=(H+H_{\rm cor})/v_{\rm pkt}$, where its velocity is
\begin{equation}
v_{\rm pkt}=c\left[1-\left(1+\frac{K_{\rm up}}{m_{\rm pkt}c^2}\right)^{-2}\right]^{1/2}.
\label{eq9}
\end{equation}
For a group of events, let $N_{{\rm arr},b}$ be the number of arrivals and $N_{{\rm trig},b}$ the number satisfying the reconnection condition. We write $N_{{\rm trig},b}=N_{{\rm arr},b}$ when $\Theta_b\geq\Theta_{\rm crit}$, and $N_{{\rm trig},b}=0$ otherwise. The total numbers are
\begin{equation}
N_{\rm arr}=\sum_b N_{{\rm arr},b},\qquad N_{\rm trig}=\sum_b N_{{\rm trig},b}.
\label{eq10}
\end{equation}
In our calculations, $N_{\rm trig}/N_{\rm arr}=1.000$. Thus, almost all modeled arrivals satisfy the adopted reconnection condition. The differences among the models mainly arise from the number of events and the subsequent release of magnetic energy, rather than from changes in the triggering fraction.

We assume that the magnetic structures activated by different events in an annulus are randomly distributed in azimuth. Their combined effective area is
\begin{equation}
A_{{\rm union},b}=A_{\rm ring}\left[1-\left(1-\frac{A_{{\rm arc},b}}{A_{\rm ring}}\right)^{N_{{\rm trig},b}}\right].
\label{eq11}
\end{equation}
When only a few events occur, the combined area is approximately proportional to the number of events. When events become frequent, different reconnection regions overlap, and the total area approaches $A_{\rm ring}$. Thus the magnetic energy available for reconnection does not increase without limit when many events occur at the same time.

For each group of triggered events, the energy transferred from the stored coronal magnetic field to the reconnecting regions is
\begin{equation}
E_{B,\rm com}=U_B\frac{A_{\rm union}}{A_{\rm ring}}.
\label{eq12}
\end{equation}
This energy is subtracted from $U_B$. To describe the effective reconnection area associated with each arrival, we define
\begin{equation}
N_{\rm arc,eq}=\sum_{b:E_{B,\rm com,b}>0}\frac{A_{{\rm union},b}}{A_{{\rm arc},b}}.
\label{eq13}
\end{equation}
This quantity gives the total effective area in units of the area of one reconnection event. It is not the actual number of independent magnetic loops. We use $N_{\rm arc,eq}/N_{\rm arr}$ to measure how the effective reconnection area per arrival changes as the event rate increases.

The magnetic energy transferred to the reconnecting regions is not released instantaneously. We use $E_{\rm arc}$ for the energy in coronal reconnection regions that has not yet been dissipated. Its evolution is

\begin{equation}
E_{\rm arc}(t+\Delta t)=e^{-\Delta t/t_{\rm rec}}\left[E_{\rm arc}(t)+E_{B,\rm com}(t)\right],
\label{eq14}
\end{equation}
the magnetic energy released during the timestep is
\begin{equation}
E_{B,\rm rel}(t)=\left[E_{\rm arc}(t)+E_{B,\rm com}(t)\right]-E_{\rm arc}(t+\Delta t),
\label{eq15}
\end{equation}
thus $U_B$ is the magnetic energy still stored in the corona, $E_{\rm arc}$ is the energy already transferred to current sheets but not yet released, and $E_{B,\rm rel}$ is the magnetic energy actually dissipated during the current timestep. The ordinary coronal input continuously replenishes $U_B$, while incoming plasma can reduce it by triggering reconnection. The balance between replenishment and repeated reconnection determines how much magnetic energy is available for later events.

\subsection{X-Ray and UV Emission}
\label{subsec:arcade_reconnection}

The X-ray energy input has two components. One is the ordinary coronal dissipation, $E_{\rm bg,diss}$. The other includes the energy dissipated in collisions with the incoming plasma, $E_{\rm sh}$, and the magnetic energy released by triggered reconnection, $E_{B,\rm rel}$. Therefore, the energies injected into the X-ray calculation during a timestep are $E_{\rm X,bg,inj}=\eta_{\rm X}E_{\rm bg,diss}$ and $E_{\rm X,sBH,inj}=\eta_{\rm X}(E_{\rm sh}+E_{B,\rm rel})$, respectively.

We calculate soft X-rays in the 0.3--2~keV band and hard X-rays in the 2--10~keV band. We assume a photon spectrum $dN/dE\propto E^{-\Gamma}$ with $\Gamma=1.9$ \citep[e.g.,][]{2017ApJS..233...17R}. The fraction of the 0.3--10~keV energy in band $j$ is $f_j=\int_{E_{j,\min}}^{E_{j,\max}}E^{1-\Gamma}\,dE/\int_{0.3}^{10}E^{1-\Gamma}\,dE$. We obtain $f_{\rm soft}=0.497$ and $f_{\rm hard}=0.503$. The same band fractions are used for the ordinary and sBH-triggered components.

We do not assume that all the dissipated energy appears immediately as observable X-ray emission. Instead, we use a finite response time to describe the production and escape of X-ray photons. For emission component $a$ and energy band $j$, the X-ray energy evolves as
\begin{equation}
\frac{dE_{a,j}}{dt}
=\dot E_{X,a,j,\mathrm{in}}-\frac{E_{a,j}}{t_{X,j}},
\label{eq16}
\end{equation}
here $a\in\{{\rm bg},{\rm sBH}\}$ denotes the ordinary and sBH-related emission components, while $j\in\{{\rm soft},{\rm hard}\}$ denotes the two X-ray bands. The quantity $E_{a,j}$ is the energy stored in the corresponding radiative component, and $\dot E_{X,a,j,\mathrm{in}}$ is the input power. The response time is $t_{{\rm X},j}=H_{\rm cor}(1+\tau_{\rm T})/c+N_{{\rm sc},j}H_{\rm cor}/(c\tau_{\rm T})$, which includes photon escape and delays due to thermal Compton scattering. We assume the same response time for the two emission components. The mean scattering number is $N_{{\rm sc},j}=\max[0,\ln(E_{{\rm repr},j}/E_{\rm seed})/\ln A_C]$, where $E_{\rm seed}=10$~eV is the seed-photon energy. We use $E_{{\rm repr},j}=1$ and 5~keV for the soft and hard bands, respectively. The mean energy amplification per scattering is $A_C=1+4\theta_e+16\theta_e^2$, where $\theta_e=kT_e/(m_ec^2)$ and $kT_e=100$~keV \citep[e.g.,][]{2015MNRAS.451.4375F,2018MNRAS.480.1819R}. The luminosity in each band is $L_{{\rm X},a,j}=E_{a,j}/t_{{\rm X},j}$. By summing the soft and hard bands, we obtain the ordinary coronal luminosity $L_{\rm X,bg}$ and the sBH-related luminosity $L_{\rm X,sBH}$. The total luminosity is $L_{\rm X}=L_{\rm X,bg}+L_{\rm X,sBH}$.

We define the fraction of the total X-ray energy released through the sBH-triggered component as
\begin{equation}
f_{X,\rm trig}=\frac{\int L_{\rm X,sBH}(t)\,dt}{\int[L_{\rm X,bg}(t)+L_{\rm X,sBH}(t)]\,dt},
\label{eq17}
\end{equation}
the integrals cover the full simulation interval used for the energy calculation. We note that $f_{X,\rm trig}$ is not the fraction of X-ray energy supplied directly by the sBHs. The energy $E_{B,\rm rel}$ may come from magnetic energy that was already stored in the corona before the plasma arrived. Similarly, $\eta_{\rm X}$ gives the fraction of dissipated energy assigned to the selected X-ray band. It does not imply that all remaining energy stays in the coronal magnetic field.

\begin{figure*}[t]
\centering
\includegraphics[width=0.8\textwidth]{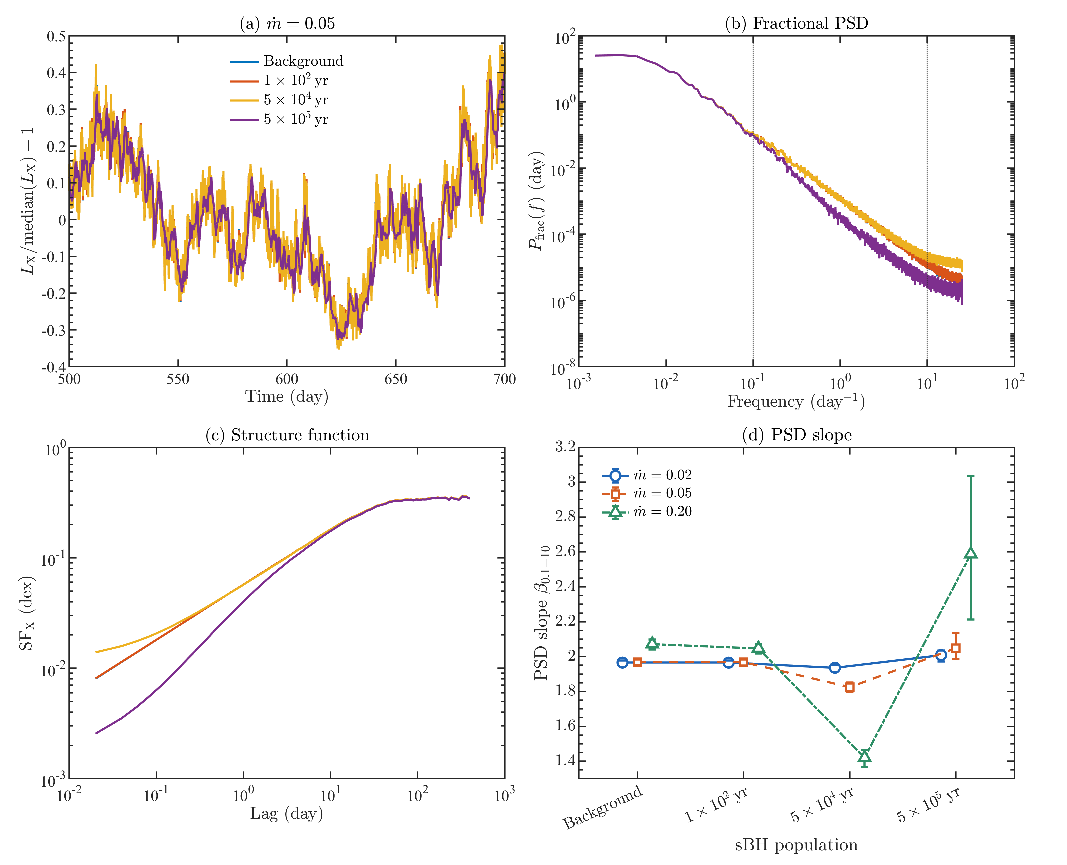}
\caption{X-ray variability for different populations of embedded sBHs. Panels (a)--(c) show the X-ray light curves normalized by their medians, the fractional PSDs, and the X-ray SFs, respectively, at $\dot m=0.05$. The fractional PSD is calculated from the normalized luminosity fluctuations $x(t)=[L_{\rm X}(t)-\langle L_{\rm X}\rangle]/\langle L_{\rm X}\rangle$, so that the integrated power gives the fractional variance. Panel (d) compares the PSD slopes $\beta_{0.1-10}$ fitted over $0.1$--$10~{\rm day}^{-1}$ for three accretion rates (the fitting interval is indicated by vertical dashed lines in panel b). The background model contains no embedded sBHs, while the other models correspond to population ages of $t_{\rm pop}=10^2$, $5\times10^4$, and $5\times10^5$ yr. The PSDs, SFs, and fitted slopes are summarized by their medians over 25 realizations. Error bars in panel (d) indicate the 16th--84th percentile ranges.}
\label{fig2}
\end{figure*}

We calculate the UV emission at 1928~\AA, corresponding to the effective wavelength of the Swift/UVOT UVW2 filter, to facilitate comparison with existing AGN monitoring observations. The time-dependent viscous heating rate $Q_{\rm vis}(R,t)$ and magnetic-coupling heating rate $Q_{\rm mc}(R,t)$ use the same radially correlated stochastic fluctuations as the coronal calculation. The ordinary disk heating after gas depletion is $Q_{\rm bg,UV}(R,t)=f_{\rm gap}[Q_{\rm vis}(R,t)+Q_{\rm mc}(R,t)]$. The energy from disk reconnection that does not escape the disk also heats the surrounding gas. For annulus $R$, let $E_{{\rm I},n}(R)=\sum_{k\in n}{\cal A}_{{\rm I},k}E_{\rm I}(R)$ be the total event energy during timestep $n$. The energy retained in the disk is converted into local heating over a finite time:
\begin{equation}
\begin{split}
Q_{{\rm sBH,UV}}^{n}(R)=& ~e^{-\Delta t/\tau_{\rm d}}Q_{{\rm sBH,UV}}^{n-1}(R)+
\\
&\frac{(1-\eta_{\rm up})E_{{\rm I},n}(R)}{A_{\rm ring}(R)}\frac{1-e^{-\Delta t/\tau_{\rm d}}}{\Delta t},
\end{split}
\label{eq18}
\end{equation}
here $\tau_{\rm d}=\max(t_{\rm I},\Delta t)$. The response is normalized so that integration over area and time gives the energy retained in the disk. Thus each set of reconnection events heats the disk through $(1-\eta_{\rm up})E_{{\rm I},n}$ and supplies kinetic energy to the corona through $\eta_{\rm up}E_{{\rm I},n}$. We use the same event sequence for both contributions.

The total disk heating rate is $Q_{\rm heat,UV}=Q_{\rm bg,UV}+Q_{\rm sBH,UV}$. We assign 30\% of this heating to a rapidly radiating disk surface component and 15\% to another rapidly responding surface layer. The adopted heating fractions are phenomenological parameters of the present UV emission calculation. The remaining 55\% heats the deeper disk gas. The deep-layer heating rate is $Q_{\rm th}=0.55Q_{\rm heat,UV}$, and the temperature evolves as
\begin{equation}
\frac{d\ln T_{\rm c,UV}}{dt}=\frac{Q_{\rm th}-Q_{\rm rad}}{\Pi C_\beta},
\label{eq19}
\end{equation}
here $Q_{\rm rad}$ is the local radiative cooling rate, and $\Pi$ is the vertically integrated pressure. The heat-capacity factor is
\begin{equation}
C_\beta=12(1-\beta_{\rm gas})
+\frac{\beta_{\rm gas}}{\gamma_{\rm gas}-1}
+\frac{(4-3\beta_{\rm gas})^2}{1+\beta_{\rm gas}},
\label{eq20}
\end{equation}
where $\beta_{\rm gas}=\Pi_{\rm gas}/\Pi$ and $\gamma_{\rm gas}=5/3$. We adopt this thermal-response prescription from Paper~I and the corona-heated accretion-disk reprocessing (CHAR) model \citep[e.g.,][]{2020ApJ...891..178S,2020ApJ...892...63C,2026ApJ..1010..134X}. Gas depletion changes not only $\Sigma$, but also the thermal quantities in the calculation, including $\Pi$, $H$, the optical depth, and $Q_{\rm rad}$.

The total 1928~\AA\ emission includes radiation from the deep disk gas, the rapidly responding surface layers, and X-ray irradiation. The irradiation heating term is $C_{\rm irr}L_{\rm X}(t-R/c)/(4\pi R^2)$, where $C_{\rm irr}=0.025$. We include the light-travel delay $R/c$ and use $\max(H_{\rm UV}/R,10^{-4})$ for the geometric factor of the incident radiation. Here $H_{\rm UV}$ is the height of the UV-emitting layer \citep[e.g.,][]{2007MNRAS.380..669C,2016MNRAS.462L..56L}. Finally, we integrate the blackbody specific intensity over both sides of the disk to obtain the monochromatic luminosity.

\subsection{Light Curves and Variability Analysis}
\label{subsec:timing_methods}

\begin{table*}[t]
\centering
\caption{Paired X-ray variability measurements and UV--X-ray lags.}
\label{tab:timing}
\resizebox{0.8\textwidth}{!}{%
\begin{tabular}{ccccccc}
\toprule
$\dot m$ & $t_{\rm pop}$ (yr) & $\Delta\beta$ &
${\cal R}_{\rm SF}(0.1\,{\rm d})$ & ${\cal R}_{\rm SF}(1\,{\rm d})$ &
$\Delta\tau_{\rm total}$ (d) & $\Delta\tau_{\rm cor}$ (d)\\
\midrule

0.02 & $10^{2}$ & $-0.000061^{+0.000005}_{-0.000034}$ & $1.000122^{+0.000005}_{-0.000003}$ & $1.000010^{+0.000004}_{-0.000001}$ & $-2.07^{+0.05}_{-0.05}$ & $\approx0$ \\
0.05 & $10^{2}$ & $-0.00046^{+0.00002}_{-0.00008}$ & $1.00092^{+0.00006}_{-0.00003}$ & $1.000079^{+0.000015}_{-0.000005}$ & $-2.41^{+0.07}_{-0.06}$ & $\approx0$ \\
0.20 & $10^{2}$ & $-0.026^{+0.001}_{-0.001}$ & $1.0344^{+0.0007}_{-0.0008}$ & $1.0031^{+0.0001}_{-0.0001}$ & $-0.22^{+0.04}_{-0.02}$ & $\approx0$ \\
0.02 & $5\times10^{4}$ & $-0.026^{+0.001}_{-0.001}$ & $1.0475^{+0.0004}_{-0.0009}$ & $1.0033^{+0.0002}_{-0.0003}$ & $-2.07^{+0.05}_{-0.05}$ & $-0.00012^{+0.00006}_{-0.00003}$ \\
0.05 & $5\times10^{4}$ & $-0.140^{+0.002}_{-0.003}$ & $1.1487^{+0.0040}_{-0.0005}$ & $0.9957^{+0.0012}_{-0.0014}$ & $-2.43^{+0.09}_{-0.08}$ & $-0.0022^{+0.0003}_{-0.0005}$ \\
0.20 & $5\times10^{4}$ & $-0.663^{+0.021}_{-0.008}$ & $0.9831^{+0.0046}_{-0.0055}$ & $0.7513^{+0.0072}_{-0.0060}$ & $-0.69^{+0.18}_{-0.05}$ & $-0.45^{+0.07}_{-0.04}$ \\
0.02 & $5\times10^{5}$ & $0.043^{+0.001}_{-0.002}$ & $0.4920^{+0.0100}_{-0.0042}$ & $0.7763^{+0.0035}_{-0.0033}$ & $-2.47^{+0.01}_{-0.05}$ & $-0.397^{+0.006}_{-0.003}$ \\
0.05 & $5\times10^{5}$ & $0.084^{+0.013}_{-0.013}$ & $0.3533^{+0.0115}_{-0.0065}$ & $0.6996^{+0.0029}_{-0.0017}$ & $-2.91^{+0.07}_{-0.06}$ & $-0.600^{+0.003}_{-0.004}$ \\
0.20 & $5\times10^{5}$ & $0.512^{+0.253}_{-0.077}$ & $0.2115^{+0.0001}_{-0.0009}$ & $0.5565^{+0.0018}_{-0.0011}$ & $-1.29^{+0.39}_{-0.20}$ & $-1.14^{+0.42}_{-0.01}$ \\

\bottomrule
\end{tabular}}
\par\smallskip
\begin{minipage}{\textwidth}\footnotesize
Note: Values are medians of 25 paired stochastic realizations. The uncertainties are 68\% bootstrap confidence intervals for the medians. We define $\Delta\beta=\beta_{\rm full}-\beta_{\rm bg}$, ${\cal R}_{\rm SF}={\rm SF}_{\rm full}/{\rm SF}_{\rm bg}$, $\Delta\tau_{\rm total}=\tau_{\rm full}-\tau_{\rm bg}$, and $\Delta\tau_{\rm cor}=\tau_{\rm full}-\tau_{\rm disk-only}$. A positive UV--X-ray lag indicates that UV variations follow the X-rays. For entries marked $\approx0$, the median satisfies $|\Delta\tau_{\rm cor}|<10^{-4}$~d, and the uncertainty range is also small.
\end{minipage}
\end{table*}

We perform 25 random realizations for each accretion rate and population age. Each simulation covers 2000~days with a timestep $\Delta t=0.02$~day. The first 500~days are excluded from the variability analysis. At fixed accretion rate, we use the same random sequence for different population ages and comparison models. This allows us to distinguish the effects of the sBH population from the underlying stochastic variability.

We use four calculations to compare the UV--X-ray lags. The full model includes the sBH population, disk heating, plasma collisions in the corona, and triggered magnetic reconnection. The disk-only model retains the same sBH population, gas depletion, and disk heating, but excludes the additional coronal-triggered dissipation. Its X-ray light curve uses the same photon-response treatment. The deep-layer-only model uses the full X-ray light curve but includes only the UV emission from the deep disk gas, without the rapidly responding surface emission or X-ray irradiation. The reference model contains no embedded sBHs. Comparing the full and disk-only models therefore allows us to estimate the additional effect of coronal reconnection on the lag while keeping the same sBH population and disk heating.

We calculate the X-ray PSDs from the normalized fluctuations $x(t)=[L_{\rm X}(t)-\langle L_{\rm X}\rangle]/\langle L_{\rm X}\rangle$. After removing a linear trend, we use Welch averaging with 400-day Hamming windows and 50\% overlap \citep[e.g.,][]{1967ITAE...15...70W,2003MNRAS.345.1271V}. We fit $P(f)\propto f^{-\beta}$ over $0.1$--$10~{\rm day}^{-1}$ and define $\Delta\beta=\beta_{\rm full}-\beta_{\rm bg}$ to describe the change in the PSD slope relative to the reference model.

We define the logarithmic X-ray structure function as ${\rm SF}_{\rm X}(\tau)=\langle[\log_{10}L_{\rm X}(t+\tau)-\log_{10}L_{\rm X}(t)]^2\rangle^{1/2}$ \citep[e.g.,][]{1985ApJ...296...46S,2003MNRAS.345.1271V}. We calculate the SF at $\tau=0.1$, 1, 10, and 100~days. We also define ${\cal R}_{\rm SF}(\tau)={\rm SF}_{\rm full}(\tau)/{\rm SF}_{\rm bg}(\tau)$. A value above unity indicates enhanced variability relative to the reference model at that timescale, while a value below unity indicates reduced variability.

The UV--X-ray lag is measured from the centroid of the continuous part of the main cross-correlation-function (CCF) peak above $0.8\,{\rm CCF}_{\max}$ \citep[e.g.,][]{1998PASP..110..660P}. A positive lag means that UV variations follow the X-rays. For each paired realization, we define $\Delta\tau_{\rm total}=\tau_{\rm full}-\tau_{\rm bg}$ and $\Delta\tau_{\rm cor}=\tau_{\rm full}-\tau_{\rm disk-only}$. The latter gives the additional lag associated with triggered coronal reconnection in the adopted model comparison. Table~\ref{tab:timing} lists the median values and 68\% bootstrap confidence intervals of the medians. Table~\ref{tab:physical} gives the median values of the magnetic-energy quantities. The uncertainties of $f_{\rm X,trig}$ and $f_{\rm res}$ in that table are the 16th--84th percentile ranges of the 25 realizations.

\section{Results}
\label{sec:results}

\subsection{X-Ray Light Curves and Variability}
\label{subsec:xray_timing_results}

Figure~\ref{fig2} shows the X-ray light curves, PSDs, and SFs for different sBH populations. At $t_{\rm pop}=10^2$~yr, the results are close to those of the reference model without embedded sBHs. At $5\times10^4$~yr, the 0.1-day SFs for $\dot m=0.02$ and 0.05 are higher than the reference values, with ratios of 1.0475 and 1.1487, respectively. However, at $\dot m=0.20$, the ratio is $0.9831^{+0.0046}_{-0.0055}$, indicating a slight reduction in the 0.1-day variability. The fitted PSD slope changes by $\Delta\beta=-0.663^{+0.021}_{-0.008}$. Thus, a flatter fitted PSD does not necessarily imply a larger SF at a particular short timescale.

At $t_{\rm pop}=5\times10^5$~yr, the 0.1-day SF is lower than the reference value for all three accretion rates. The ratios are 0.492, 0.353, and 0.212 for $\dot m=0.02$, 0.05, and 0.20, respectively. The changes in the 1-day SF are smaller. This shows that the effects of embedded sBHs on the X-ray emission are most evident at short timescales. The light curves for $\dot m=0.05$ also show different variability patterns at the intermediate and late population ages.

\begin{table*}[t]
\centering
\caption{sBH-triggered X-ray emission and coronal magnetic-energy measurements.}
\label{tab:physical}
\resizebox{0.6\textwidth}{!}{%
\begin{tabular}{cccccc}
\toprule
$\dot m$ & $t_{\rm pop}$ (yr) & $f_{X,\rm trig}$ &
$N_{\rm arc,eq}/N_{\rm arr}$ & $f_{\rm res}$ & $\chi_{\rm dep}$ \\
\midrule

0.02 & $10^{2}$ & $0.0003844^{+0.0000009}_{-0.0000005}$ & $1.000$ & $1.000$ & $0.00009$ \\
0.05 & $10^{2}$ & $0.0007601^{+0.0000025}_{-0.0000022}$ & $1.000$ & $1.000$ & $0.00017$ \\
0.20 & $10^{2}$ & $0.002812^{+0.000025}_{-0.000014}$ & $1.000$ & $1.000$ & $0.00063$ \\
0.02 & $5\times10^{4}$ & $0.20165^{+0.00007}_{-0.00025}$ & $0.971$ & $0.99955^{+0.00031}_{-0.00106}$ & $0.045$ \\
0.05 & $5\times10^{4}$ & $0.3523^{+0.0009}_{-0.0016}$ & $0.948$ & $0.9872^{+0.0024}_{-0.0036}$ & $0.090$ \\
0.20 & $5\times10^{4}$ & $0.7923^{+0.0133}_{-0.0132}$ & $0.825$ & $0.734^{+0.012}_{-0.010}$ & $0.426$ \\
0.02 & $5\times10^{5}$ & $0.9128^{+0.0140}_{-0.0090}$ & $0.786$ & $0.596^{+0.009}_{-0.006}$ & $0.605$ \\
0.05 & $5\times10^{5}$ & $0.9630^{+0.0144}_{-0.0122}$ & $0.721$ & $0.459^{+0.007}_{-0.006}$ & $0.758$ \\
0.20 & $5\times10^{5}$ & $0.9978^{+0.0015}_{-0.0036}$ & $0.312$ & $0.2665^{+0.0004}_{-0.0010}$ & $0.960$ \\

\bottomrule
\end{tabular}}
\par\smallskip
\begin{minipage}{\textwidth}\footnotesize
Note: The quantity $f_{X,\rm trig}$ is the fraction of X-ray energy released through sBH-triggered dissipation. The ratio $N_{\rm arc,eq}/N_{\rm arr}$ gives the effective reconnection area per arrival in units of the area of one event. The filling fraction $f_{\rm res}$ is evaluated before magnetic energy is transferred to a reconnection region when plasma arrives. The global mean deficit $\chi_{\rm dep}$ is evaluated after energy transfer. Values are medians of 25 stochastic realizations. The uncertainties of $f_{X,\rm trig}$ and $f_{\rm res}$ are the 16th--84th percentile ranges. Only median values are given for the other two quantities. The early-age values of $f_{\rm res}$ are approximately unity at the quoted precision.
\end{minipage}
\end{table*}

\subsection{Effects of Coronal Magnetic Reconnection on X-Ray Variability}
\label{subsec:physical_origin_results}

To investigate the change in short-timescale variability, we compare the available coronal magnetic energy and its release. Table~\ref{tab:physical} gives the fraction of X-ray energy emitted through sBH-triggered dissipation, the effective reconnection area per arrival, and the filling and deficit fractions of the coronal magnetic energy.

At $\dot m=0.20$, $f_{X,\rm trig}$ increases from 0.792 at $5\times10^4$~yr to 0.998 at $5\times10^5$~yr, while the 0.1-day SF ratio decreases from 0.983 to 0.212. Over the same interval, the effective reconnection area per arrival, $N_{\rm arc,eq}/N_{\rm arr}$, decreases from 0.825 to 0.312. The magnetic-energy filling fraction $f_{\rm res}$ decreases from 0.734 to 0.2665, while the global energy deficit $\chi_{\rm dep}$ increases from 0.426 to 0.960. Thus, although more X-ray energy is released through sBH-triggered dissipation, less coronal magnetic energy is available before each event.

The total fractional rms variability does not decrease by the same amount. For $\dot m=0.20$, we define
\begin{equation}
F_{\rm var}\equiv\frac{{\rm std}(L_{\rm X})}{\langle L_{\rm X}\rangle}.
\label{eq21}
\end{equation}
The reference model has $F_{\rm var}=0.597$, while the model at $t_{\rm pop}=5\times10^5$~yr has $F_{\rm var}=0.584$. At the same time, its 0.1-day SF is only 0.212 times the reference value. The change is therefore mainly seen at shorter timescales. In the adopted model, a larger sBH population produces more disturbances in the corona, but the reconnection regions overlap more strongly and less magnetic energy remains available for subsequent events. Both effects are consistent with the decrease in short-timescale variability.

\subsection{UV--X-Ray Time Lags}
\label{subsec:uvx_results}

\begin{figure*}[t]
\centering
\includegraphics[width=0.9\textwidth]{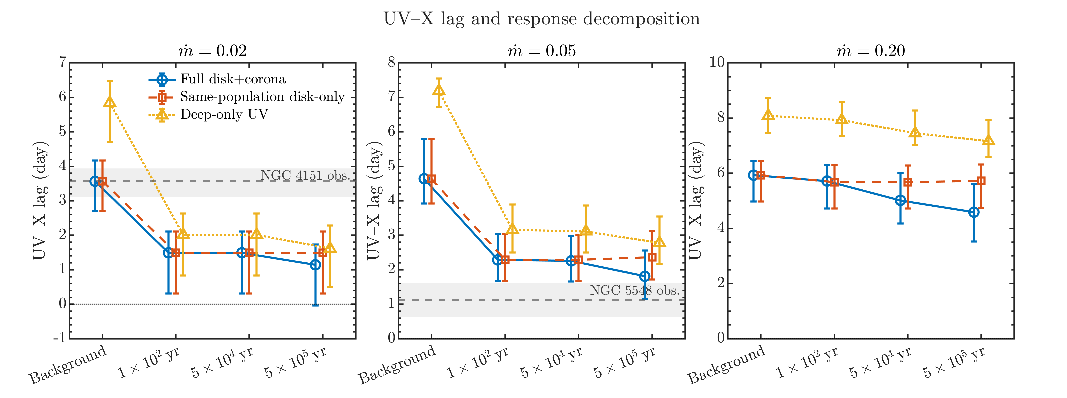}
\caption{UV--X-ray lags for different model components. Blue circles show the full disk-corona model, red squares show the disk-only model with the same sBH population, and yellow triangles show the deep-layer UV response paired with the full X-ray light curve. The gray shaded regions show the observed X-ray--UVW2 lags of NGC~4151, $3.58^{+0.36}_{-0.46}$~days, and NGC~5548, $1.12\pm0.49$~days, in the $\dot m=0.02$ and 0.05 panels, respectively \citep{2017ApJ...840...41E,2018MNRAS.480.2881M}. These measurements are shown only for comparison of the lag scales; no sBH population age is assigned to either source. Error bars indicate the 16th--84th percentile ranges from 25 realizations.}
\label{fig3}
\end{figure*}

Figure~\ref{fig3} shows the UV--X-ray lags calculated for the different disk and coronal emission components. At $\dot m=0.02$ and 0.05, the lags of the full and disk-only models are very similar at the early and intermediate population ages. This indicates that the lag changes in these cases mainly arise from the disk response, including gas depletion and heating by magnetic reconnection in the disk, rather than from additional reconnection in the corona.

At early population ages, the X-ray variability remains close to that of the reference model, while the UV--X-ray lag is already shorter. As shown in Paper I, embedded sBHs can affect the thermal response of the AGN disk and the UV emission from its inner regions. These changes may lead to a faster UV response, even when the additional X-ray emission associated with the sBHs is still weak. Therefore, the early decrease in the UV--X-ray lag is mainly associated with changes in the disk emission rather than with the additional coronal reconnection.

The effect of triggered coronal reconnection on the lag becomes more evident at the late age and high accretion rate. At $\dot m=0.20$ and $t_{\rm pop}=5\times10^5$~yr, the difference between the full and disk-only models reaches $\Delta\tau_{\rm cor}=-1.14$~day. In contrast, the largest increase in the 0.1-day SF relative to the reference occurs at $\dot m=0.05$ and $t_{\rm pop}=5\times10^4$~yr. Thus, the model with the strongest increase in short-timescale X-ray variability is not the one with the largest additional lag from the corona.

For comparison, Figure~\ref{fig3} also shows the observed X-ray--UVW2 lags of NGC~4151 and NGC~5548 \citep{2017ApJ...840...41E,2018MNRAS.480.2881M}. The calculated lags have similar orders of magnitude to the observations. However, we do not assign a particular sBH population age to either AGN. The observed X-ray bands also differ from the 0.3--10~keV band used here. Therefore, these observations are included only as references for the lag timescales.

\section{Discussion}
\label{sec:discussion}

\subsection{Physical Interpretation}
\label{subsec:timing_window}

Our results show that the X-ray variability associated with embedded sBHs depends on both the rate of disk reconnection events and the magnetic energy available in the corona. At intermediate population ages and moderate accretion rates, much of the coronal magnetic energy remains stored, and the reconnection regions associated with different events overlap only moderately. The additional magnetic-energy release can therefore increase the short-timescale X-ray variability.

At later population ages, more plasma enters the corona and magnetic reconnection events become more frequent. However, the reconnection regions overlap more strongly. At the same time, the stored magnetic energy decreases because energy is repeatedly released. Thus, even when a larger fraction of the X-ray emission comes from the sBH-associated coronal dissipation, the short-timescale flux variations need not be stronger. At $\dot m=0.20$, this behavior already appears at the intermediate population age.

The UV--X-ray lags also provide information about the disk and corona. Gas depletion and local disk heating can change the UV light curves even when the additional coronal X-ray variability is weak. At high accretion rates and late population ages, the extra lag associated with coronal reconnection becomes more significant. Observations also show that the correlation between AGN X-ray and UV/optical emission can differ from the correlations among UV/optical bands \citep[e.g.,][]{2015ApJ...806..129E,2017ApJ...835...65S,2017MNRAS.470.3591G,2017ApJ...840...41E,2018MNRAS.480.2881M}. The X-ray SF and UV--X-ray lag therefore provide different information about the emission.

\subsection{Observational Implications}
\label{subsec:observational_implications}

The sBH population age and the coronal magnetic-energy filling fraction cannot be measured directly from observed light curves. The X-ray PSD, SF, and UV--X-ray lag are more suitable quantities for comparison with observations. Our model predicts only weak additional X-ray variability when there are few sBHs in the disk. At intermediate population ages, the short-timescale variability can either increase or decrease, depending on the accretion rate. At late ages, it decreases for all three accretion rates considered here. Since AGN X-ray variability also depends on the SMBH mass and accretion rate, comparisons should use reference samples with similar values of these parameters \citep[e.g.,][]{2006Natur.444..730M,2012A&A...544A..80G}.

For the adopted $M_\bullet=5\times10^7\,M_\odot$, the largest differences occur on timescales shorter than one day. Measuring the 0.1-day SF requires sampling substantially finer than 0.1~day and a sufficiently long observing baseline. Intensive \textit{Swift} monitoring of NGC~4151 achieved sampling of about 6~hr over approximately 69~days. NGC~5548 was monitored with subday sampling over about 125~days \citep[e.g.,][]{2015ApJ...806..129E,2017ApJ...840...41E}. These observations show that intensive monitoring is possible. However, 6~hr sampling alone cannot resolve variability on a 0.1-day timescale. Denser observations or multiple monitoring campaigns may be needed. Simultaneous UV observations would also help distinguish changes in the disk thermal response from variations in the X-ray corona. Since embedded sBH populations may differ among AGNs, a large sample of X-ray light curves is needed to investigate their possible timing signatures. Future X-ray timing surveys could provide useful samples for comparing the short-timescale variability of AGNs with similar SMBH masses and accretion rates.

The optical/UV variability discussed in Paper~I may help identify AGNs with substantial populations of embedded objects. The X-ray timing properties of these candidates can then be compared with those of AGNs matched in SMBH mass and accretion rate. Combining the short-timescale SF with the UV--X-ray lag may provide a better test of the model than searching for strong X-ray flares alone.

\subsection{Model Limitations}
\label{subsec:limitations}

We adopt prescribed disk and coronal structures rather than performing self-consistent radiation-magnetohydrodynamic simulations. We consider only $M_\bullet=5\times10^7\,M_\odot$ and $\alpha=0.2$, with fixed prescriptions for the coronal magnetic field and reconnection. The event-amplitude width $\sigma_{\rm I}$, coronal-to-disk height ratio $H_{\rm cor}/H$, upward energy fraction $\eta_{\rm up}$, and X-ray radiation fraction $\eta_{\rm X}$ are also fixed. The UV calculation covers only 1928~\AA. We do not include reflection, absorption, or self-consistent spectral evolution.

Changing these assumptions would affect the event rates, replenishment of coronal magnetic energy, and X-ray response. A different SMBH mass or viscosity parameter would also change the disk structure and the embedded sBH population. Therefore, our results cannot be applied to other AGNs by simply rescaling the timescales. The calculations show that frequent magnetic reconnection can change X-ray variability within the adopted model. They do not determine the true magnetic-field geometry of an individual AGN corona or provide a general observational detection threshold.

\section{Conclusions}
\label{sec:conclusions}

Using the sBH populations and disk reconnection events calculated in Paper~I, we have studied their effects on AGN X-ray variability. Our model includes the kinetic energy carried by plasma into the corona, the replenishment and release of coronal magnetic energy, and the X-ray and UV responses. The main conclusions are as follows.

\begin{enumerate}
\item The short-timescale X-ray variability depends on the sBH population age and accretion rate. At $\dot m=0.05$, the 0.1-day SF is 1.149 times the reference value at $5\times10^4$~yr, but decreases to 0.353 times the reference value at $5\times10^5$~yr. At $\dot m=0.20$, the 0.1-day SF is already slightly reduced at the intermediate age and decreases to 0.212 times the reference value at the late age.

\item A larger fraction of X-ray emission arising from sBH-associated coronal dissipation does not necessarily imply stronger short-timescale variability. As the number of events increases, the reconnection regions overlap more strongly. The replenishment of stored coronal magnetic energy also becomes insufficient to offset its repeated release. Both effects contribute to the decrease in the short-timescale SF predicted by the model.

\item The UV--X-ray lags are affected by both the disk and coronal responses. Changes in disk heating can alter the lag even when the additional coronal emission is weak. At late ages and high accretion rates, the contribution of triggered coronal reconnection to the lag becomes more significant.

\item The predicted variability depends on the adopted magnetic reconnection and energy-replenishment prescriptions. Testing these results requires intensive subday X-ray monitoring, simultaneous UV observations, and suitable reference samples.
\end{enumerate}

\section*{Acknowledgments}
We thank Bao-Quan Huang for helpful discussions. This work was supported by the National Key R\&D Program of China under grant 2023YFA1607902, the National Natural Science Foundation of China under grants 12494572, 12221003, 125B2058, 12373070, 12192223, 12494575, and 12273005. During the process of writing this article, we utilized ChatGPT (from OpenAI) to assist with language polishing and to help structure the discussion. The authors remain responsible for all scientiﬁc assumptions, analyses, and interpretations.

\end{document}